\documentclass{article}
\usepackage[english]{babel}
\usepackage{amsfonts,amsmath,amssymb}
\usepackage{bm}
\usepackage[a4paper]{geometry}
\usepackage{mathtools} 
\usepackage{setspace}
\usepackage{standalone}
\usepackage{subfig}
\usepackage{tikz}
\usepackage{hyperref}
\usetikzlibrary{quotes}
\usetikzlibrary{angles, arrows.meta, calc, decorations.markings,
  decorations.pathmorphing, fadings, patterns, quotes, intersections}

\doublespace

\hypersetup{
  pdfborder = {0 0 0.1}
}

\newcommand{\p}{\partial}

\newcommand{\R}{\mathbb{R}}
\newcommand{\Z}{\mathbb{Z}}

\providecommand{\pd}[3][]{\ensuremath{
  \frac{\p^{#1}#2}{\p#3^{#1}}
}}
\providecommand{\pdi}[3][]{\ensuremath{
  \p^{#1}#2/\p #3^{#1}
}}

\DeclarePairedDelimiter{\dprod}{\langle}{\rangle}
\newcommand{\defeq}{\coloneqq} 
\DeclarePairedDelimiter{\abs}{\lvert}{\rvert} 
\DeclarePairedDelimiter{\norm}{\lVert}{\rVert} 

\title{Computational study of the dynamics of an asymmetric wedge billiard}
\author{K.D.
Anderson\footnote{\href{mailto:kdanderson@uj.ac.za}{kdanderson@uj.ac.za}} \qquad
C.M. Villet\footnote{\href{mailto:cmvillet@uj.ac.za}{cmvillet@uj.ac.za}} \\
Department of Mathematics and Applied Mathematics \\
University of Johannesburg
}
\date{}

\begin{document}
\maketitle
\begin{abstract}
  We introduce the asymmetric wedge billiard as a generalization of the wedge
  billiard first introduced and studied by Lehtihet and Miller in 1986.
  This is a billiard system in which the billiard ball moves under the influence
  of a constant gravitational field, colliding elastically with two wedge walls
  with the collisions obeying the reflection law. Collision maps are given from
  which derivatives and area-preservation (or lack thereof) were determined.
  Expressions for the fixed points of the collision maps were also calculated
  and discussed.
  Long-term dynamics were determined computationally from which we observed
  integrable, quasi-periodic and chaotic behaviour which were all dependent on
  the wedge angles.
\end{abstract}

\maketitle 

\section{Introduction}
A dynamical billiard system consists of a particle represented as a geometric
point moving freely within a bounded region in the plane, its collisions with
the boundary of the region are elastic and obey the reflection law.

G.D. Birkhoff \cite{birkhoff1927periodic} introduced dynamical billiards as a
means to prove Poincar\'e's last geometric conjecture.
Others \cite{berry1981regularity,lazutkin1973existence,poritsky1950billiard}
continued his work on convex billiards with some open questions remaining to
this day.
The seminal work by Y.G. Sinai \cite{sinai1970dynamical} introduced a new class
of billiards, called dispersing billiards, as an application to modelling
Lorentz gas and was the first to show that these billiard systems are chaotic.
Another class of billiards, i.e., polygonal billiards, arose naturally from the
study of another mechanical system, that of two point particles moving on a
straight line between two walls.
This shows the utility of dynamical billiards, as Birkhoff himself stated that
most Hamiltonian systems with two degrees of freedom could be studied by the
appropriate transform to a dynamical billiard.
Standard billiard dynamics are quite rich and numerous open problems remain.

Research has also been done on modifications of classical billiard
systems.
It would be natural to consider the particle moving in the quantum realm
\cite{bruus1994quantum,szeredi1993classical2,waalkens1997elliptic}
or moving relativistically
\cite{deryabin2003generalized1,deryabin2003generalized2,deryabin2004exponential}.
Other billiard systems consider modifications to the region of motion itself,
for example, a hole or multiple holes within the region---these are the
so-called ``open billiards''; billiard systems where the boundary changes in
time
\cite{kamphorst1999bounded,koiller1995time,ladeira2008scaling,lenz2007classical,lenz2007scattering,lenz2009evolutionary};
and billiard systems where the billiard moves under the influence of a constant
force field, either magnetic
\cite{berglund1996integrability,da2000periodic,gongora2002classical,robnik1985classical,tasnadi1996behavior}
or gravitational \cite{da2015circular,korsch1991new,lehtihet1986numerical}.

\begin{figure}
  \begin{center}
      \begin{tikzpicture}[scale=2,>=stealth, baseline=0]
    \coordinate (o) at (0,0);
    \draw[thick] (-1.5,0) coordinate (-x) -- (1.5,0) coordinate (x);
    \draw[dashed] (0,0) -- (0,2) coordinate (y);
    \draw[thick] (o) -- (1.25,1.5) coordinate (a);
    \draw[thick] (o) -- (-1.25,1.5) coordinate (b);
    \path (a) -- (o) -- (y) pic ["$\theta$", draw, angle radius=5mm, angle
    eccentricity=1.5] {angle=a--o--y};
    \path (b) -- (o) -- (y) pic ["$\theta$", draw, angle radius=5mm, angle
    eccentricity=1.5] {angle=y--o--b};
    \fill[pattern=north east lines] (o) -- (b) -- ($(o)!(b)!(-x)$);
    \fill[pattern=north east lines] (o) -- (a) -- ($(o)!(a)!(x)$);
    \draw[->] (-.8,2) -- node [midway, right] {$m\bm{g}$} (-.8,1.75);
    \coordinate (c0) at ($(o)!0.9!(a)$);
    \coordinate (c1) at ($(o)!0.5!(b)$);
    \coordinate (c2) at ($(o)!0.3!(a)$);
    \draw (c0) parabola bend (0.35, 2) (c1);
    \draw (c1) parabola bend (-0.3, 0.8) (c2);
  \end{tikzpicture}
    \caption{The (symmetric) wedge billiard.}
    \label{fig:wedge-billiard}
  \end{center}
\end{figure}
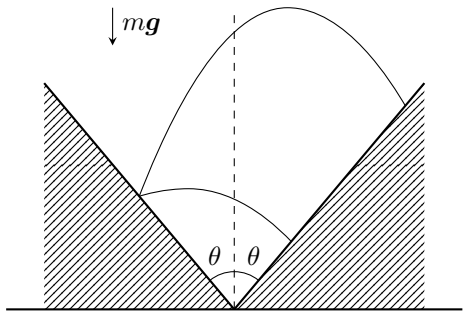

The wedge billiard (illustrated in Figure \ref{fig:wedge-billiard}) is a
billiard system where the particle moves within a constant
gravitational force field, it was first studied by Lehtihet and Miller
\cite{lehtihet1986numerical}.
They showed that the dynamics of the billiard was dependent on the wedge angle
$\theta$.
Richter, Scholz, and Wittek \cite{richter1990breathing} classified the symmetric
periodic orbits of the wedge billiard using symmetry lines
\cite{birkhoff1927dynamical,greene1981universal,pina1987symmetry} which lead to
the description of the so-called ``breathing chaos''---the regular variation
between chaotic and quasi-periodic behaviour for certain parameter values of the
wedge.
Szeredi
\cite{szeredi1993classical,szeredi1996hard,szeredi1992periodic,szeredi1993classical1,szeredi1993classical2}
studied the wedge billiard in the quantum context whilst
Korsch and Lang \cite{korsch1991new} modified the wedge billiard by changing the
shape of the boundary to a parabola and found that the dynamics are integrable.
Hartl, Miller, and Mazzoleni \cite{hartl2013dynamics} studied the dynamics of
various gravitational billiards, including the wedge billiard, with boundaries
which were driven sinusoidally.
The wedge billiard has found some applications in engineering and physics.
Sepulchre and Gerard \cite{sepulchre2003stabilization} applied the wedge
billiard model with some modification to stabilize an elementary impact control
system which applications in robotics, whilst Choi, Sundaram and Raizen
\cite{choi2010single} applied the wedge billiard model to the problem of
single-photon cooling.

One of the main assumptions of the wedge billiard is that the wedge is symmetric
with respect to the vertical axis as seen in Figure \ref{fig:wedge-billiard}.
We considered the case of the \emph{asymmetric wedge} in which no assumptions
were made about the wedge angle(s).
There are only two references
\cite{lehtihet1986numerical,wojtkowski1990system} about the asymmetric wedge
billiard in the literature.
Lehtihet and Miller \cite{lehtihet1986numerical} mentions the asymmetric wedge
in the context of their self-gravitating system with three different mass
densities.
Their assumption that lead to the wedge billiard were that the mass densities
were similar while unequal mass densities would result in an asymmetric wedge
billiard.
Wojtkowski \cite{wojtkowski1990system} studied a system of one-dimensional balls
under the influence of gravity to illustrate his principles
\cite{wojtkowski1986principles} for the design of billiards with nonvanishing
Lyapunov exponents.
Wojtkowski then provided a transformation between the system and the asymmetric
wedge and established that the asymmetric wedge billiard will have nonvanishing
Lyapunov exponents for $\theta_1 + \theta_2 > \pi/2$.

The purpose of this paper is to further the study of some of the dynamics of the
asymmetric wedge billiard.

\section{Model}
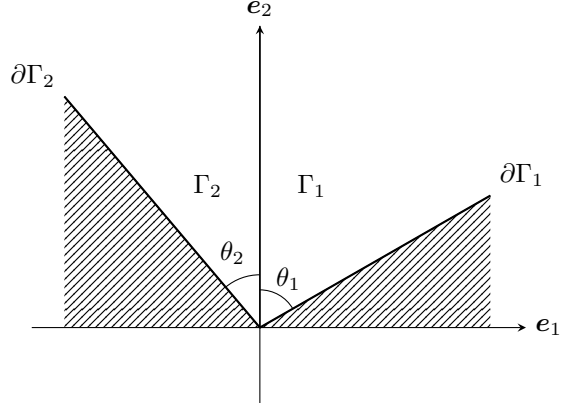
\begin{figure}
  \begin{center}
      \begin{tikzpicture}[scale=2,>=stealth]
    \coordinate (o) at (0,0);
    \draw[->] (-1.5,0) coordinate (-x) -- (1.75,0) coordinate (x);
    \node at (x) [right] {$\bm{e}_1$};
    \draw[->] (0,-.5) -- (0,2) coordinate (y);
    \node at (y) [above] {$\bm{e}_2$};
    \draw[thick] (o) -- ++(30:1.75) coordinate (a);
    \node at (a) [above right] {$\partial\Gamma_1$};
    \draw[thick] (o) -- ++(130:2) coordinate (b);
    \node at (b) [above left] {$\partial\Gamma_2$};
    \draw (a) -- (o) -- (y) pic ["$\theta_1$", draw, angle radius=5mm, angle
    eccentricity=1.5] {angle=a--o--y};
    \draw (b) -- (o) -- (y) pic ["$\theta_2$", draw, angle radius=7mm, angle
    eccentricity=1.5] {angle=y--o--b};
    \node at (70:1) [circle,fill=white,inner sep=1pt] {$\Gamma_1$};
    \node at (110:1) [circle,fill=white,inner sep=1pt] {$\Gamma_2$};
    \fill[pattern=north east lines] (o) -- (b) -- ($(-x)!(b)!(o)$);
    \fill[pattern=north east lines] (o) -- (a) -- ($(x)!(a)!(o)$);
  \end{tikzpicture}
  \end{center}
  \caption{Geometry of the asymmetric wedge billiard.}
  \label{fig:awb-particle-rom}
\end{figure}

Consider the two planar regions defined as
\begin{subequations}\label{eqn:awb-particle-allowed-region}
\begin{align}
  \mathcal{Q}_1 &= \left\{ (x, y) \in \R^2 : x \geq 0,\; y > x\cot(\theta_1)
  \right\}, \label{eqn:awb-particle-allowed-region-rhs} \\
  \mathcal{Q}_2 &= \left\{ (x, y) \in \R^2 : x < 0,\; y > -x\cot(\theta_2)
  \right\} \label{eqn:awb-particle-allowed-region-lhs}
\end{align}
\end{subequations}
with respective boundaries defined as
\begin{subequations}\label{eqn:awb-particle-allowed-region-boundary}
\begin{align}
  \p \mathcal{Q}_1 &= \left\{ (x, y) \in \R^2 : x \geq 0,\; y =
  x\cot(\theta_1) \right\}, \label{eqn:awb-particle-allowed-region-boundary-rhs}
  \\
  \p \mathcal{Q}_2 &= \left\{ (x, y) \in \R^2 : x < 0,\; y =
  -x\cot(\theta_2) \right\}.
  \label{eqn:awb-particle-allowed-region-boundary-lhs}
\end{align}
\end{subequations}
Here $\R^2$ is a normed space with inner product $\dprod{\bm{x},
\bm{y}}$ and induced norm $\norm{\bm{x}} = \sqrt{\dprod{\bm{x},\bm{x}}}$, where
$\bm{x}, \bm{y} \in \R^2$.
We define the standard basis of $\R^2$ as $\mathcal{B}_s \defeq \{\bm{e}_1,
\bm{e}_2\}$ which correspond to the horizontal and vertical references axes
illustrated in Figure \ref{fig:awb-particle-rom}.
The angles $\theta_1$ and $\theta_2$ are respectively measured clockwise and
anticlockwise from the reference axis $\bm{e}_2$ to the straight lines
representing $\p \mathcal{Q}_1$ and $\p \mathcal{Q}_2$ as illustrated in Figure
\ref{fig:awb-particle-rom}.

We consider the motion of a point particle of mass $m$ within a (constant)
gravitational field $\bm{g}$ within the region $\bar{\mathcal{Q}} \defeq
\bar{\mathcal{Q}}_1 \cup \bar{\mathcal{Q}}_2$, where $\bar{\mathcal{Q}}_j \defeq
\mathcal{Q}_j \cup \p \mathcal{Q}_j$ ($j \in \{1,2\}$).
We shall call $\bar{\mathcal{Q}}$ the \emph{allowed region of motion} for the
particle.
We shall refer to the set $\p \mathcal{Q} \defeq \p \mathcal{Q}_1 \cup \p
\mathcal{Q}_2$ as the \emph{asymmetric wedge}; when $\theta_1 = \theta_2$ we
shall call $\p \mathcal{Q}$ the \emph{symmetric wedge}.
The boundaries $\p \mathcal{Q}_j$, $j=\left\{1, 2\right\}$, are referred to as
\emph{wedge walls}; the line $\p \mathcal{Q}_1$ ($\p \mathcal{Q}_2$
respectively) is called the \emph{right-hand wall} (\emph{left-hand wall}
respectively).
The intersection of $\p \mathcal{Q}_1$ and $\p \mathcal{Q}_2$ is called the
\emph{wedge vertex}.

Respectively, let $\bm{q} \defeq \bm{q}(t) \in \bar{\mathcal{Q}}$ be the
position vector, $\bm{p} \defeq \bm{p}(t) \in \R^2$ be the momentum vector (such
that $\bm{p}^2 = \dprod{\bm{p},\bm{p}} = 1$), and $E \in \R^+$ be the
mechanical energy of the particle.
If we fix an angle $\phi$ with respect to the fixed basis vector $\bm{e}_1$,
then we may rewrite $\bm{p}$ as $\bm{p} = (\cos(\phi), \sin(\phi)) \in
\mathbb{S}^1$ where $\mathbb{S}^1 = \{\bm{x} \in \mathbb{R}^2 : \lVert \bm{x}
\rVert = 1\}$.
The phase space of the particle may be described by the set
\begin{equation}\label{eqn:awb-full-phase-space}
  \mathcal{P} \defeq \bar{\mathcal{Q}} \times \mathbb{S}^1 = \left\{ (\bm{q},
  \bm{p}) : \bm{q} \in \bar{\mathcal{Q}}, \; \bm{p} \in \mathbb{S}^1 \right\}
\end{equation}
together with the projection mappings $\pi_{\bm{q}} : \mathcal{P} \to
\bar{\mathcal{Q}}$, $\pi_{\bm{p}} : \mathcal{P} \to \mathbb{S}^1$ such that
$\pi_{\bm{q}}(\bm{x}) = \bm{q}$ and $\pi_{\bm{p}}(\bm{x}) = \bm{p}$, where
$\bm{x} = (\bm{q}, \bm{p})$.
On this phase space we may define the energy function (or Hamilton function)
$H : \mathcal{P} \to \R$ such that
\begin{equation}\label{eqn:energy-function-general}
  H(\bm{q}, \bm{p}) = \frac{\bm{p}^2}{2} + U(\bm{q})
\end{equation}
where $U$ is a scalar potential satisfying $\pdi{U}{\bm{q}} = -\bm{g}$.
The energy function is independent of time and hence it is constant along
solution curves, therefore we may set $H(\bm{q}, \bm{p}) = E$.

\begin{figure}
  \begin{center}
    \begin{tikzpicture}[scale=2,>=stealth]
    \coordinate (o) at (0,0);
    \draw[->] (-.1,0) -- (1,0) coordinate (x);
    \draw[->] (0,-.1) -- (0,2.25) coordinate (y);
    \node at (x) [right] {$\bm{e}_1$};
    \node at (y) [above] {$\bm{e}_2$};
    \draw[thick] (o) -- ++(40:2) coordinate (w1);
    \draw[thick] (o) -- ++(120:2) coordinate (w2);
    \draw (w1) -- (o) -- (y) pic [draw, angle radius=7mm] {angle=w1--o--y};
    \node at (75:0.5) {$\theta_1$};
    \draw (y) -- (o) -- (w2) pic [draw, "$\theta_2$", angle eccentricity=1.5]
    {angle=y--o--w2};
    \draw[densely dotted] ($(o)!0.8!(w1)$) parabola bend (.5,1.5) ($(o)!0.4!(w2)$);
    \draw[->] (o) -- node[pos=0.7, sloped, above] {$\bm{q}(t)$} ++(60:1.6325) coordinate (q);
    \path[->] (x) -- (o) -- (q) pic [draw] {angle=x--o--q};
    \draw[->] (q) -- ($(o)!1.25!(q)$) coordinate (e1);
    \draw[->] (q) -- ($(q)!1.25!90:(e1)$) coordinate (e2);
    \filldraw (q) circle [radius=0.3pt];
    \node at (e1) [right] {$\bar{\bm{e}}(t)_1$};
    \node at (e2) [above] {$\bar{\bm{e}}(t)_2$};
    \draw[rotate=60] (q) rectangle ++(.05,.05);
    \node at (0:7pt) [below] {$\varphi(t)$};
  \end{tikzpicture}
  \caption{Reference frames used in the study of the asymmetric wedge billiard.}
  \label{fig:awb-reference-system2}
  \end{center}
\end{figure}
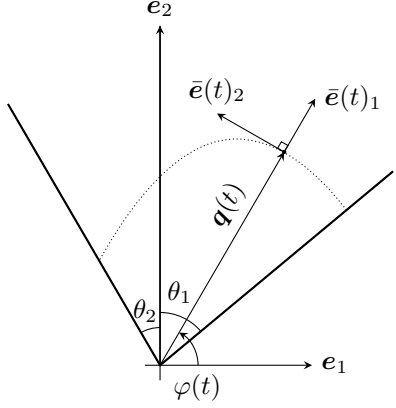

By careful transformation \cite{anderson2019thesis} the vector components and
the energy become dimensionless quantities such that $m = g= E = 1$, which we
shall assume throughout the rest of the article.
We shall let $x$ and $y$ denote the components of $\bm{q}$ with respect to
$\bm{e}_1$ and $\bm{e}_2$ and, similarly, we denote by $u$ and $w$ the
components of $\bm{p}$ with respect to $\bm{e}_1$ and $\bm{e}_2$.
We shall also make use of a secondary reference system, as illustrated by Figure
\ref{fig:awb-reference-system2}, with basis vectors $\mathcal{B}_r =
\{\bm{\bar{e}}_1, \bm{\bar{e}}_2\}$.
Transformation between the two bases is accomplished through a rotation by the
angle $\varphi \defeq \varphi(t)$ measured from $\bm{e}_1$ to the position
vector $\bm{q}(t)$, i.e.
\begin{equation}\label{eqn:awb-basis-transformation}
  \begin{bmatrix}
    \bm{\bar{e}}_1 \\ \bm{\bar{e}}_2
  \end{bmatrix}
  =
  \mathsf{R}(\varphi)
  \begin{bmatrix}
    \bm{e}_1 \\ \bm{e}_2
  \end{bmatrix}, \quad
  \mathsf{R}(\varphi) \defeq
  \begin{bmatrix}
    \cos(\varphi) & -\sin(\varphi) \\
    \sin(\varphi) & \cos(\varphi)
  \end{bmatrix}
\end{equation}
We denote by $\bar{u} \defeq p\cos(\phi - \varphi)$ and $\bar{w} \defeq
p\sin(\phi - \varphi)$ the components of $\bm{p}$ with respect to the
$\mathcal{B}_r$ basis; it follows that we may consider $\bm{p} \in \mathbb{S}^1$
with angle parameter $\phi - \varphi$ in this instance.
From the transformation \eqref{eqn:awb-basis-transformation} we obtain
\begin{equation}\label{eqn:awb-momentum-component-transformation}
  \begin{bmatrix}
    \bar{u} \\ \bar{w}
  \end{bmatrix}
  =
  \begin{bmatrix}
    \cos(\varphi) & -\sin(\varphi) \\
    \sin(\varphi) & \cos(\varphi)
  \end{bmatrix}
  \begin{bmatrix}
    u \\ w
  \end{bmatrix}
\end{equation}
which relates the components of $\bm{p}$ in the $\mathcal{B}_s$ and
$\mathcal{B}_r$ bases to each other.
In terms of the $x, y, u, w$ coordinates the energy function becomes
\begin{equation}\label{eqn:energy-function-standard-basis}
  H(x,y,u,w) = \frac{u^2 + w^2}{2} + y
\end{equation}
and in the $x,y,\bar{u},\bar{w}$ coordinates the energy function becomes
\begin{equation}\label{eqn:energy-function-radial-basis}
  H(x,y,\bar{u},\bar{w}) = \frac{\bar{u}^2 + \bar{w}^2}{2} + y.
\end{equation}

\subsection{Collision maps}
It can be shown from first principles \cite{anderson2019thesis} by solving the
Hamilton equations of motion derived from \eqref{eqn:energy-function-general}
that the particle moves along a parabolic path between collisions with the wedge
walls.
Collisions are elastic due to energy conservation; these collisions obey the
law of reflection, that is, the angle of incidence equals the angle of
reflection (the standard assumption for billiard systems).
We assume any other type of dissipation is completely absent from the system.
We also assume that the particle will keep moving until such time that it
collides with the wedge vertex at which point the motion will stop.
Thus the time interval of the motion can either be finite (a collision with the
vertex) or infinite (no collision with the vertex at all) depending on the
initial conditions.

Furthermore, the $x$ and $y$ variables are related by the straight line
equations describing $\p \mathcal{Q}_1$ and $\p \mathcal{Q}_2$.
The value of the $y$ variable can easily be determined from
\eqref{eqn:energy-function-standard-basis} or
\eqref{eqn:energy-function-radial-basis}.
Hence the only variables that need to be determined at collisions are
the momentum components $u$, $w$ or $\bar{u}$, $\bar{w}$.
We keep to the convention established \cite{lehtihet1986numerical} and make use
of the coordinates $\bar{u}$, $\bar{w}$ in the $\mathcal{B}_r$ basis.

For successive collisions on $\p \mathcal{Q}_1$ we define the map $F_A : \p
\mathcal{Q}_1 \to \p \mathcal{Q}_1$, $(\bar{u}_j, \bar{w}_j^2) \mapsto (\bar{u}_{j+1},
\bar{w}_{j+1}^2)$ with
\begin{equation}\label{eqn:awb-collision-map-rhs-rhs}
  \bar{u}_{j+1} = \bar{u}_j - 2\bar{w}_j\cot(\theta_1), \quad \bar{w}_{j+1}^2 =
  \bar{w}_j^2.
\end{equation}
For a collision between the particle, starting from $\p \mathcal{Q}_1$, with
$\p \mathcal{Q}_2$ we define the map $F_B : \p \mathcal{Q}_1 \to \p \mathcal{Q}_2$,
$(\bar{u}_j, \bar{w}_j^2) \mapsto (\bar{u}_{j+1}, \bar{w}_{j+1}^2)$ with
\begin{equation}\label{eqn:awb-collision-map-rhs-lhs}
  \begin{aligned}
    \bar{u}_{j+1} &= \frac{\bar{w}_j\cos(\theta_1) - \bar{w}_{j+1}\cos(\theta_2) -
    \bar{u}_j\sin(\theta_1)}{\sin(\theta_2)}, \\
    \bar{w}_{j+1}^2 &= \frac{2\sin(\theta_2)\sin(\theta_1 +
    \theta_2)}{\cos(\theta_1)}\left(1 - \frac{\bar{u}_j^2 +
    \bar{w}_j^2}{2}\right) + (\bar{u}_j\sin(\theta_1 + \theta_2) +
    \bar{w}_j\cos(\theta_1 + \theta_2))^2.
  \end{aligned}
\end{equation}
Setting $\theta_1 = \theta_2 = \theta$ in \eqref{eqn:awb-collision-map-rhs-rhs}
and \eqref{eqn:awb-collision-map-rhs-lhs} and simplifying results in the maps
for the symmetric wedge billiard
\cite{lehtihet1986numerical,richter1990breathing}.

Similarly, for successive collisions on $\p \mathcal{Q}_2$ we define the map
$G_A : \p \mathcal{Q}_2 \to \p \mathcal{Q}_2$, $(\bar{u}_j, \bar{w}_j^2) \mapsto
(\bar{u}_{j+1}, \bar{w}_{j+1}^2)$ with
\begin{equation}\label{eqn:awb-collision-map-lhs-lhs}
  \bar{u}_{j+1} = \bar{u}_j + 2\bar{w}_j\cot(\theta_2), \quad \bar{w}_{j+1}^2 =
  \bar{w}_j^2.
\end{equation}
For a collision between the particle, starting from $\p \mathcal{Q}_2$, with $\p
\mathcal{Q}_1$ we define the map $G_B : \p \mathcal{Q}_2 \to \p \mathcal{Q}_1$,
$(\bar{u}_j, \bar{w}_j^2) \mapsto (\bar{u}_{j+1}, \bar{w}_{j+1}^2)$ with
\begin{equation}\label{eqn:awb-collision-map-lhs-rhs}
  \begin{aligned}
    \bar{u}_{j+1} &= \frac{-\bar{w}_j\cos(\theta_2) -
    \bar{w}_{j+1}\cos(\theta_1) - \bar{u}_j\sin(\theta_2)}{\sin(\theta_1)}, \\
    \bar{w}_{j+1}^2 &= \frac{2\sin(\theta_1)\sin(\theta_1 +
    \theta_2)}{\cos(\theta_2)}\left(1 - \frac{\bar{u}_j^2 +
    \bar{w}_j^2}{2}\right) + (\bar{u}_j\sin(\theta_1 + \theta_2) +
    \bar{w}_j\cos(\theta_1 + \theta_2))^2.
  \end{aligned}
\end{equation}
We note that the maps \eqref{eqn:awb-collision-map-lhs-lhs} and
\eqref{eqn:awb-collision-map-lhs-rhs} can be transformed into those of the
symmetric wedge billiard by setting $\theta_1 = \theta_2$ and taking into
consideration of an appropriate substitution to factor in the symmetry about the
vertical axis.
A full derivation, from first principles, of the maps
\eqref{eqn:awb-collision-map-rhs-rhs}-\eqref{eqn:awb-collision-map-lhs-rhs}
found in the first author's thesis \cite{anderson2019thesis}.

\section{Dynamics}
\label{sec:awb-dynamics}
The choice between using $F_A$ and $F_B$ is determined from the inequality
$\left(\bar{u}_j - 2\bar{w}_j\cot(\theta_1)\right)^2 + \bar{w}_j^2 \leq 2$
which may be derived from the energy equation
\eqref{eqn:energy-function-radial-basis}.
Similarly, the choice between using $G_A$ and $G_B$ is determined from the
inequality $\left(\bar{u}_j + 2\bar{w}_j\cot(\theta_2)\right)^2 + \bar{w}_j^2
\leq 2$.
Choosing between mappings $F$ and $G$ is determined completely by the value
of horizontal component of the particle's position.

We now define the collision space $\mathcal{C} = \partial \mathcal{Q} \times
\mathbb{S}^1$.
The tuple $(\mathcal{C}, \left\{ F_A, F_B, G_A, G_B \right\})$ constitutes a
discrete dynamical system.
The orbit of collisions points is determined from compositions of the maps
\eqref{eqn:awb-collision-map-rhs-rhs}-\eqref{eqn:awb-collision-map-lhs-rhs},
that is, if $\bm{x} = (x, y, \bar{u}, \bar{w}) \in \mathcal{C}$ we determine,
for example, $F_i \circ G_j(\bm{x})$ or $G_i^k\circ F_B(\bm{x})$ where $i,j =
\left\{A,B\right\}$ and $k \in \mathbb{N}$.
However, not all combinations of compositions correspond to physically possible
collisions.
Compositions which are excluded are
\begin{align*}
  G_A &\circ F_A, & G_B &\circ F_A, & F_A &\circ G_A, & F_B &\circ G_A, \\
  F_A &\circ F_B, & G_A &\circ G_B, & F_B &\circ F_B, & G_B &\circ G_B.
\end{align*}
while compositions which correspond to physically possible collisions are
\begin{align*}
    F_A &\circ F_A, & G_A &\circ G_A, & G_A &\circ F_B, & F_B &\circ G_B, \\
    F_B &\circ F_A, & G_B &\circ G_A, & F_B &\circ G_B, & G_B &\circ F_B.
\end{align*}
Any number of combinations from this last collection may constitute the orbit
$\mathcal{O}(\bm{x}_0)$ of some initial point $\bm{x}_0 \in \mathcal{C}$.

\subsection{Derivative of the collision maps}
The derivative of a map may be used to determine if the map is area-preserving
or to linearize the map in a neighbourhood of any of its fixed points
\cite{hale1991dynamics}.
In the case of the linear maps $F_A$ and $G_A$ we have
\begin{equation} \label{eqn:awb-collision-map-same-side-derivative}
  DF_A \defeq
  \begin{bmatrix}
    1 & -2\cot(\theta_1) \\ 0 & 1
  \end{bmatrix}, \quad
  DG_A \defeq
  \begin{bmatrix}
    1 & 2\cot(\theta_2) \\ 0 & 1
  \end{bmatrix}
\end{equation}
with determinants equal to unity for both these matrices.
The derivative of $F_B$ is
\begin{equation}\label{eqn:awb-collision-map-rhs-lhs-derivative}
  DF_B \defeq
  \begin{bmatrix}
    \pdi{\bar{u}_{j+1}}{\bar{u}_j} & \pdi{\bar{u}_{j+1}}{\bar{w}_j} \\
    \pdi{\bar{w}_{j+1}}{\bar{u}_j} & \pdi{\bar{w}_{j+1}}{\bar{w}_j}
  \end{bmatrix}
\end{equation}
where
\begin{align*}
  \pd{\bar{w}_{j+1}}{\bar{u}_j} &=
  \frac{1}{\bar{w}_{j+1}}\left[\Bigl(-\frac{\sin(\theta_2)\sin(\theta_1 +
  \theta_2)}{\cos(\theta_1)} + \sin^2(\theta_1 + \theta_2)\Bigr)\bar{u}_j +
  \frac{\bar{w}_j\sin\left(2(\theta_1 + \theta_2)\right)}{2}\right], \\
  \pd{\bar{w}_{j+1}}{\bar{w}_j} &=
  \frac{1}{\bar{w}_{j+1}}\left[\Bigl(-\frac{\sin(\theta_2)\sin(\theta_1 +
  \theta_2)}{\cos(\theta_1)} + \cos^2(\theta_1 + \theta_2)\Bigr)\bar{w}_j +
  \frac{\bar{u}_j\sin\left(2(\theta_1 + \theta_2)\right)}{2}\right], \\
  \pd{\bar{u}_{j+1}}{\bar{u}_j} &= -\cot(\theta_2)\pd{\bar{w}_{j+1}}{\bar{u}_j}
  - \frac{\sin(\theta_1)}{\sin(\theta_2)}, \\
  \pd{\bar{u}_{j+1}}{\bar{w}_j} &= -\cot(\theta_2)\pd{\bar{w}_{j+1}}{\bar{w}_j}
  + \frac{\cos(\theta_1)}{\sin(\theta_2)}.
\end{align*}
The determinant of $DF_B$ is
\begin{equation}\label{eqn:awb-collision-map-rhs-lhs-derivative-determinant}
  \det\left(DF_B\right) =
   \frac{\bar{w}_j\cos(\theta_2)}{\bar{w}_{j+1}\cos(\theta_1)}
\end{equation}
Similarly, the derivative of $G_B$ is
\begin{equation}\label{eqn:awb-collision-map-lhs-rhs-derivative}
  DG_B \defeq
  \begin{bmatrix}
    \pdi{\bar{u}_{j+1}}{\bar{u}_j} & \pdi{\bar{u}_{j+1}}{\bar{w}_j} \\
    \pdi{\bar{w}_{j+1}}{\bar{u}_j} & \pdi{\bar{w}_{j+1}}{\bar{w}_j}
  \end{bmatrix}
\end{equation}
where
\begin{align*}
  \pd{\bar{w}_{j+1}}{\bar{u}_j} &=
  \frac{1}{\bar{w}_{j+1}}\left[\Bigl(-\frac{\sin(\theta_1)\sin(\theta_1 +
  \theta_2)}{\cos(\theta_2)} + \sin^2(\theta_1 + \theta_2)\Bigr)\bar{u}_j +
  \frac{\bar{w}_j\sin\left(2(\theta_1 + \theta_2)\right)}{2}\right], \\
  \pd{\bar{w}_{j+1}}{\bar{w}_j} &=
  \frac{1}{\bar{w}_{j+1}}\left[\Bigl(-\frac{\sin(\theta_1)\sin(\theta_1 +
  \theta_2)}{\cos(\theta_2)} + \cos^2(\theta_1 + \theta_2)\Bigr)\bar{w}_j +
  \frac{\bar{u}_j\sin\left(2(\theta_1 + \theta_2)\right)}{2}\right], \\
  \pd{\bar{u}_{j+1}}{\bar{u}_j} &= \cot(\theta_1)\pd{\bar{w}_{j+1}}{\bar{u}_j} -
  \frac{\sin(\theta_2)}{\sin(\theta_1)}, \\
  \pd{\bar{u}_{j+1}}{\bar{w}_j} &= \cot(\theta_1)\pd{\bar{w}_{j+1}}{\bar{w}_j} -
  \frac{\cos(\theta_2)}{\sin(\theta_1)}
\end{align*}
with determinant
\begin{equation}\label{eqn:awb-collision-map-lhs-rhs-derivative-determinant}
  \det\left(DG_B\right) =
  \frac{\bar{w}_j\cos(\theta_1)}{\bar{w}_{j+1}\cos(\theta_2)}.
\end{equation}
We note that the maps $F_B$ and $G_B$ are only area-preserving whenever
$\det\left(DF_B\right) = 1$ and $\det\left(DG_B\right) = 1$, that is,
$\bar{w}_j\cos(\theta_2)/\bar{w}_{j+1}\cos(\theta_1) = 1$ for $F_B$ and
$\bar{w}_j\cos(\theta_1)/\bar{w}_{j+1}\cos(\theta_2) = 1$ for $G_B$.
Thus the maps $F_B$ and $G_B$ are area-preserving whenever $\bar{w}_{j+1} =
\bar{w}_j$ and $\theta_2 \equiv \theta_1 + 2k\pi$, $k \in \Z$.
For any value of $k \neq 0$, we would obtain a value for $\theta_2 \notin (0 ,
\pi/2)$ irrespective of the chosen value of $\theta_1$, therefore $\theta_2 =
\theta_1$ and hence we conclude that the maps are only area-preserving at the
fixed point of the symmetric wedge billiard \cite{lehtihet1986numerical}.

\subsection{Fixed points of the collision maps}
The map $F_A$ has a family of fixed points given by
\begin{equation}\label{eqn:awb-collision-map-rhs-rhs-fp}
  (\bar{u}_*, \bar{w}_*) = (c_F, 0), \quad c_F \in \R.
\end{equation}
This corresponds, physically, to the particle sliding up or down the wall $\p
\mathcal{Q}_1$ depending on whether $c_F$ is positive or negative.
This is the same family of fixed point as derived for the symmetric wedge
billiard by Lehtihet and Miller \cite{lehtihet1986numerical} and Richter
\emph{et al} \cite{richter1990breathing}.
We note that for $c_F = 0$ we obtain $(\bar{u}_*, \bar{w}_*) = (0,0)$ which is
the wedge vertex.
The fixed point of the map $F_B$ can be shown to be
\begin{equation}\label{eqn:awb-collision-map-rhs-lhs-fp}
  \begin{aligned}
    \bar{u}_* &= \bar{w}_*\tan\left(\frac{\theta_2 -
    \theta_1}{2}\right), \\
    \bar{w}_*^2 &= \frac{2\sin(\theta_2)\sin(\theta_1 +
    \theta_2)}{\left[1 + g(\theta_1,\theta_2) - \left(f(\theta_1,
    \theta_2)\right)^2\right]\cos(\theta_1)}
  \end{aligned}
\end{equation}
where
\begin{equation}\label{eqn:helper-func-def}
  \begin{aligned}
    f(\theta_1, \theta_2) &\defeq \frac{\cos((3\theta_1 +
    \theta_2)/2)}{\cos((\theta_2 - \theta_1)/2)}, \\
    g(\theta_1, \theta_2) &\defeq \frac{\sin(\theta_2)\sin(\theta_1 +
    \theta_2)}{\cos(\theta_1)\cos^2((\theta_2 - \theta_1)/2)}.
  \end{aligned}
\end{equation}

Similarly, the family of fixed points for $G_A$ is given by
\begin{equation}\label{eqn:awb-collision-map-lhs-lhs-fp}
  (\bar{u}_*, \bar{w}_*) = (c_G,0), \quad c_G \in \R,
\end{equation}
and the fixed point of $G_B$ given by
\begin{equation}\label{eqn:awb-collision-map-lhs-rhs-fp}
  \begin{aligned}
    \bar{u}_* &= \bar{w}_*\tan\left(\frac{\theta_2 -
    \theta_1}{2}\right), \\
    \bar{w}_*^2 &= \frac{2\sin(\theta_1)\sin(\theta_1 +
    \theta_2)}{\left[1 + g(\theta_1,\theta_2) - \left(f(\theta_1,
    \theta_2)\right)^2\right]\cos(\theta_2)}
  \end{aligned}
\end{equation}
with $f$ and $g$ as given in \eqref{eqn:helper-func-def}.

We were not able to determine the stability of the family of fixed points
\eqref{eqn:awb-collision-map-rhs-rhs-fp} and
\eqref{eqn:awb-collision-map-lhs-lhs-fp} analytically, since the eigenvalues of
the matrices \eqref{eqn:awb-collision-map-same-side-derivative} are both equal
to unity.
However, we can determine stability via informal argument.
For example, if we were to choose $c_F < 0$, supposing the particle is situated
on $\p \mathcal{Q}_1$, which is a member of the family
\eqref{eqn:awb-collision-map-rhs-rhs-fp}, the particle would slide down toward
the wedge vertex at which point its motion would stop.
Hence the subset of the family \eqref{eqn:awb-collision-map-rhs-rhs-fp} is
stable in the sense that all the fixed points in this subset are attracted to
the wedge vertex.
Similarly, if we were to choose $c_F > 0$, the particle would slide up the slope
and away from the wedge vertex.
Since we assumed no dissipation at all, the particle would keep sliding up for
all eternity and hence this subset of the family
\eqref{eqn:awb-collision-map-rhs-rhs-fp} is repelled away from the wedge vertex.

\begin{figure}
  \begin{center}
    \subfloat{
      \includegraphics[scale=0.45]{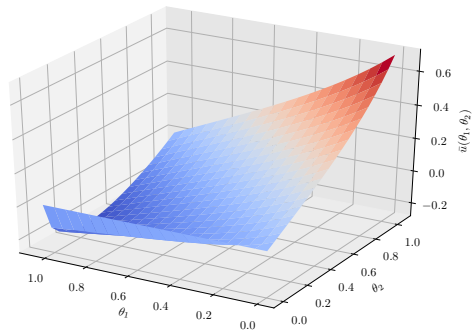}
    }
    \\
    \subfloat{
      \includegraphics[scale=0.45]{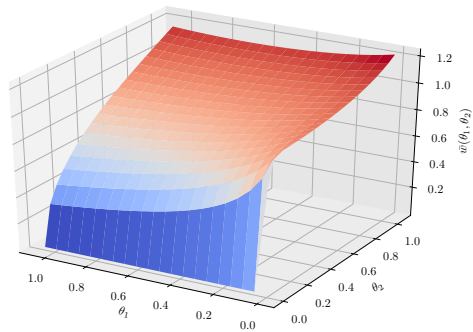}
    }
  \end{center}
  \caption{[Colour online] Fixed point ``surfaces'' for $F_B$ for various
  $\theta_1, \theta_2 \in (0, \pi/3)$.}
  \label{fig:awb-map-T3-standard-fixed-point-surface}
\end{figure}

\begin{figure}
  \begin{center}
    \subfloat{
      \includegraphics[scale=0.45]{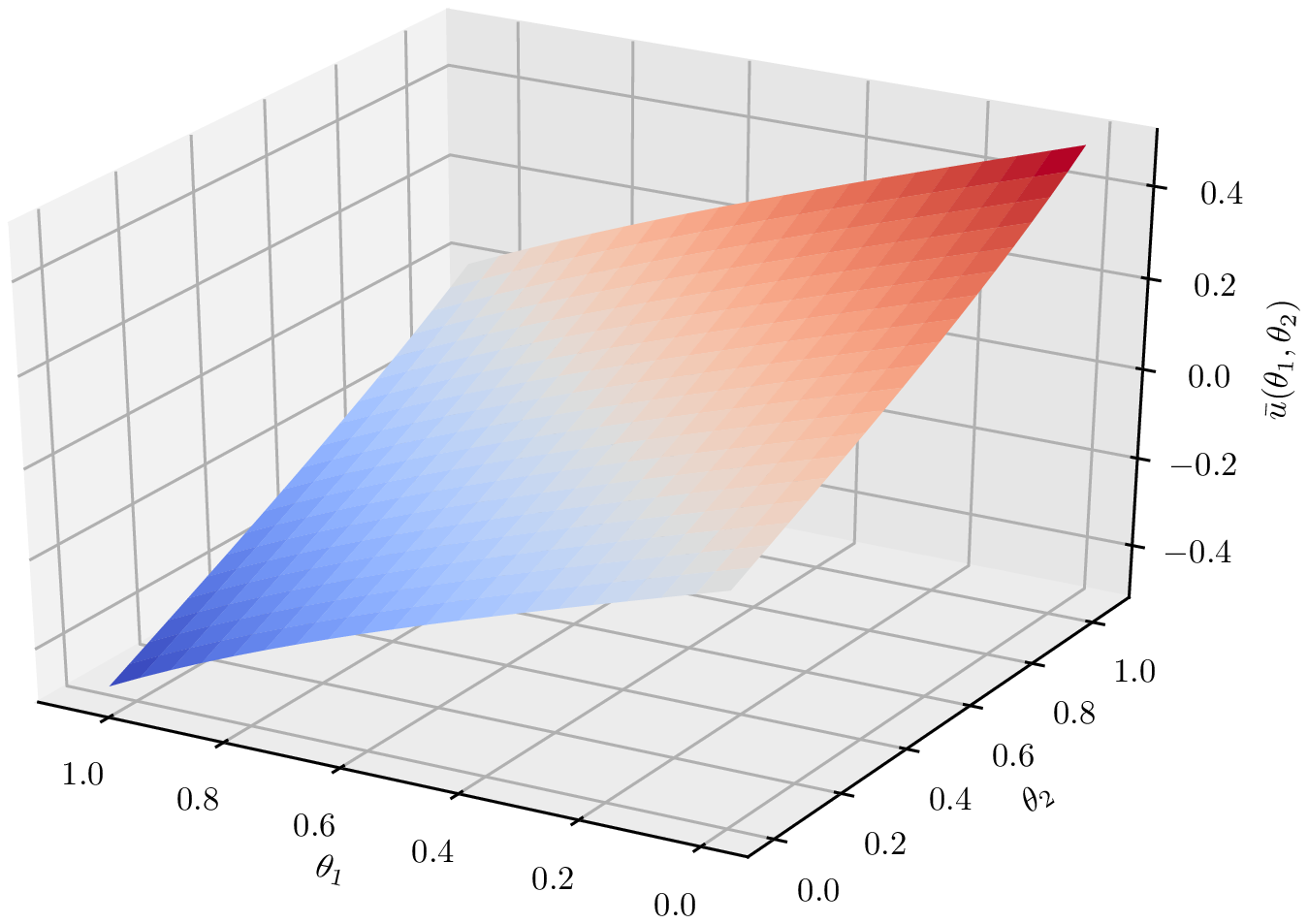}
    }
    \\
    \subfloat{
      \includegraphics[scale=0.45]{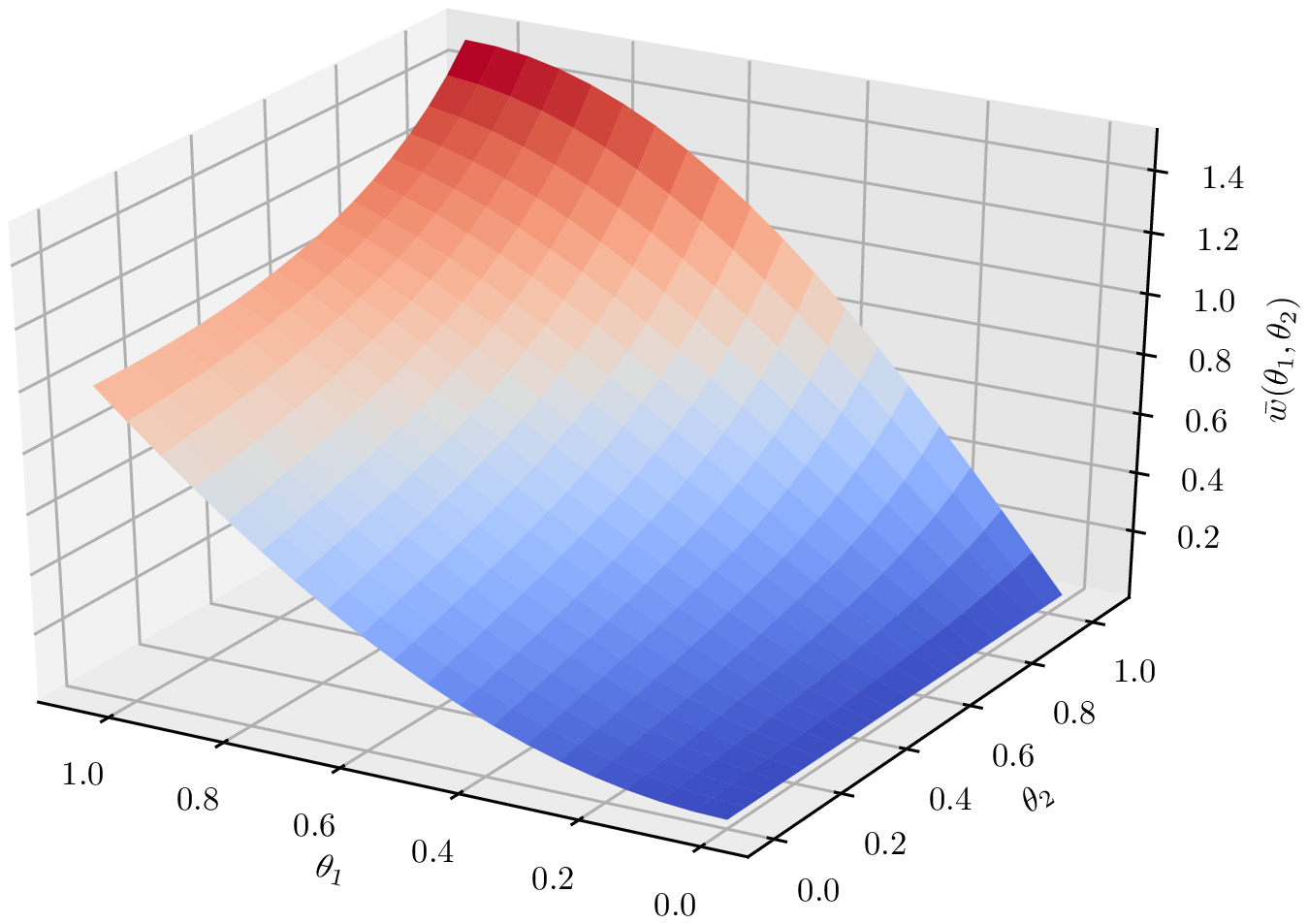}
    }
  \end{center}
  \caption{[Colour online] Fixed point ``surfaces'' for $G_B$ for various
  $\theta_1, \theta_2 \in (0, \pi/3)$.}
  \label{fig:awb-map-T4-standard-fixed-point-surface}
\end{figure}

Stability analysis of the eigenvalues of
\eqref{eqn:awb-collision-map-rhs-lhs-derivative} and
\eqref{eqn:awb-collision-map-lhs-rhs-derivative} would, of necessity, require a
numerical study and was not attempted during our original research.
However, in Figure \ref{fig:awb-map-T3-standard-fixed-point-surface} and Figure
\ref{fig:awb-map-T4-standard-fixed-point-surface} we illustrate the values
$\bar{u}_*$ and $\bar{w}_*$ take for various values of $\theta_1, \theta_2 \in
(0, \pi/3)$.
For $\theta_1 \to \pi/2$ and $\theta_2 \to \pi/2$, simultaneously, it was
observed that the ``fixed point surfaces'' nears a singularity which agrees with
the physical model---both walls would be horizontal in the limit and the motion
would be equivalent to one-dimensional motion under the influence of gravity
with elastic collisions on a horizontal surface.

\subsection{Computational Results}
\begin{figure}
  \begin{center}
    \begin{tikzpicture}[scale=2,>=Stealth]
      \draw[<->] (0,1) coordinate (y) -- (0,0) coordinate (o) -- (1,0)
      coordinate (x);
      \node at (y) [above] {$\bm{e}_2$};
      \node at (x) [right] {$\bm{e}_1$};
      \draw (o) -- ++(40:1.75) coordinate (q1);
      \draw (o) -- ++(120:1) coordinate (q2);
      \path (y) -- (o) -- (q2) pic [draw, "$\theta_2$", angle eccentricity=1.5,
      angle radius=17.5] {angle=y--o--q2};
      \path (y) -- (o) -- (q1) pic [draw, "$\theta_1$", angle eccentricity=1.5]
      {angle=q1--o--y};
      \coordinate (q0) at ($(o)!0.5!(q1)$);
      \draw[->] (q0) -- ++(60:0.75) coordinate (p0);
      \node at (q0) [below right] {$\bm{q}_0 = (x_0, y_0)$};
      \filldraw (q0) circle [radius=.2mm];
      \node at (p0) [above] {$\bm{p}_0 = (u_0, w_0)$};
      \draw[densely dashed] (q0) -- ++(0:0.5) coordinate (qx);
      \path (qx) -- (q0) -- (p0) pic [draw, "$\phi$", angle eccentricity=1.5]
      {angle=qx--q0--p0};
      \path (x) -- (o) -- (q0) pic [draw, "$\varphi$", angle eccentricity=1.5,
      angle radius=20]
      {angle=x--o--q0};
      \path (p0) -- (q0) -- (q1) pic [draw, "$\vartheta$",
      angle eccentricity=1.25, angle radius=25] {angle=q1--q0--p0};
    \end{tikzpicture}
    \caption{Graphical representation of initial conditions for computational
    simulation.}
    \label{fig:awb-launch-angle}
  \end{center}
\end{figure}
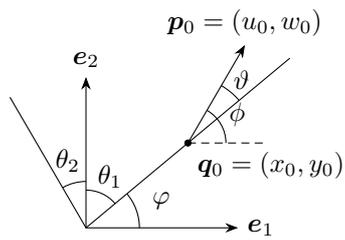

For general dynamics, we iterated the maps
\eqref{eqn:awb-collision-map-rhs-rhs}-\eqref{eqn:awb-collision-map-lhs-rhs}
for 10,000 collisions for a particle always starting on $\p \mathcal{Q}_1$.
Initial conditions were determined using an angle $\vartheta$ which is measured
anticlockwise from $\p \mathcal{Q}_1$ to the forward direction of the momentum
vector of the particle, as illustrated in Figure \ref{fig:awb-launch-angle}.
From this launch angle we then set $u_0 = -\sin(\vartheta - \theta_1)$ and
$w_0 = \cos(\vartheta - \theta_1)$, with $y_0$ determined using the energy
equation \eqref{eqn:energy-function-standard-basis}, and $x_0 =
y_0\tan(\theta_1)$; using $u_0$ and $w_0$ we then determine $\bar{u}_0$ and
$\bar{w}_0$ using the rotation transformation
\eqref{eqn:awb-momentum-component-transformation}.

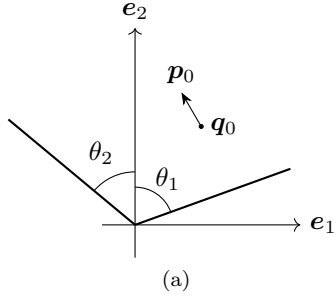
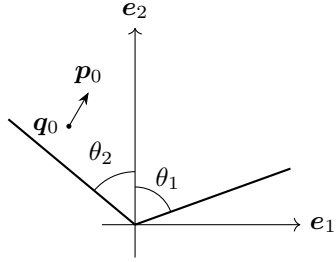
\begin{figure}
  \begin{center}
    \subfloat[\label{fig:reflection-symmetry1}]{
      \begin{tikzpicture}[baseline=0,scale=0.87]
        \coordinate (o) at (0,0);
        \draw[->] (-.5,0) -- (2.5,0) node [right] {$\bm{e}_1$};
        \draw[->] (0,-.5) -- (0,3) coordinate (y);
        \node at (y) [above] {$\bm{e}_2$};
        \draw[thick] (o) -- ++(20:2.5) coordinate (a);
        \draw[thick] (o) -- ++(140:2.5) coordinate (b);
        \path (a) -- (o) -- (y) pic [draw, "$\theta_1$", angle eccentricity=1.5]
        {angle=a--o--y};
        \path (b) -- (o) -- (y) pic [draw, "$\theta_2$", angle radius=20, angle
        eccentricity=1.5] {angle=y--o--b};
        \coordinate (q0) at (1,1.5);
        \filldraw (q0) circle [radius=.3mm];
        \node at (q0) [right] {$\bm{q}_0$};
        \draw[->,>=Stealth] (q0) -- ++(120:0.6) node [above] {$\bm{p}_0$};
      \end{tikzpicture}
    }
    \\
    \subfloat[\label{fig:reflection-symmetry3}]{
      \begin{tikzpicture}[baseline=0,scale=0.87]
        \coordinate (o) at (0,0);
        \draw[->] (-.5,0) -- (2.5,0) node [right] {$\bm{e}_1$};
        \draw[->] (0,-.5) -- (0,3) coordinate (y);
        \node at (y) [above] {$\bm{e}_2$};
        \draw[thick] (o) -- ++(20:2.5) coordinate (a);
        \draw[thick] (o) -- ++(140:2.5) coordinate (b);
        \path (a) -- (o) -- (y) pic [draw, "$\theta_1$", angle eccentricity=1.5]
        {angle=a--o--y};
        \path (b) -- (o) -- (y) pic [draw, "$\theta_2$", angle radius=20, angle
        eccentricity=1.5] {angle=y--o--b};
        \coordinate (q0) at (-1,1.5);
        \filldraw (q0) circle [radius=.3mm];
        \node at (q0) [left] {$\bm{q}_0$};
        \draw[->,>=Stealth] (q0) -- ++(60:0.6) node [above] {$\bm{p}_0$};
      \end{tikzpicture}
    }
    \caption{Reflection symmetry about the vertical axis in configuration space.
    Note that the momentum vector also needs to be reflected accordingly,
    otherwise a different orbit will be obtained.}
    \label{fig:awb-reflection-symmetry}
  \end{center}
\end{figure}

We note that there exists a reflection symmetry about the vertical axis on
condition that the particle also be reflected accordingly, as illustrated in
Figure \ref{fig:awb-reflection-symmetry} for $(\theta_1, \theta_2) = (7\pi/18,
5\pi/18)$.
This reflective symmetry corresponds to a reflection about the line $\theta_1 =
\theta_2$ in the parameter space. 
Hence we only considered parameters $\theta_1$, $\theta_2$ such that
$0 < \theta_1 < \pi/2$ and $0 < \theta_2 \leq \theta_1$.

To illustrate the dynamics observed during simulation, we plotted the results in
the dynamical system's phase space which should not be confused with the
previously defined phase space \eqref{eqn:awb-full-phase-space}.
We define the dynamical phase space as the set
\begin{equation}\label{eqn:awb-phase-space}
  \Omega \defeq \left\{ (\bar{u}, \bar{w}^2) \in \R^2 : \bar{w}^2 \geq 0, \;
  \abs{\bar{u}} \leq \sqrt{2E} \right\}.
\end{equation}
Furthermore, the parabola
\begin{equation}
  \Gamma_p \defeq \left\{ (\bar{u}, \bar{w}^2) \in \Omega : \bar{w}^2 > 0, \;
  \bar{u}^2 + \bar{w}^2 - 2E = 0 \right\}
  \label{eqn:awb-phase-space-vertex-collisions}
\end{equation}
forms the upper boundary on the phase space with the lower boundary given by
\begin{equation}\label{eqn:awb-phase-space-fp}
  \Gamma_\ell \defeq \left\{ (\bar{u}, \bar{w}^2) \in \Omega : \bar{w}^2 = 0, \;
  \abs{\bar{u}} \leq \sqrt{2E} \right\}.
\end{equation}
The area inclosed by $\p \Omega \defeq \Gamma_p \cup \Gamma_\ell$ defines the
set of allowed values that $\bar{u}$ and $\bar{w}$ may take during the
particle's motion.
Points on the parabola $\Gamma_p$ corresponds to vertex collisions while points
on the straight line $\Gamma_\ell$ corresponds to the particle sliding up or
down the wedge walls.
\begin{figure}
  \begin{center}
    \includegraphics[scale=0.6]{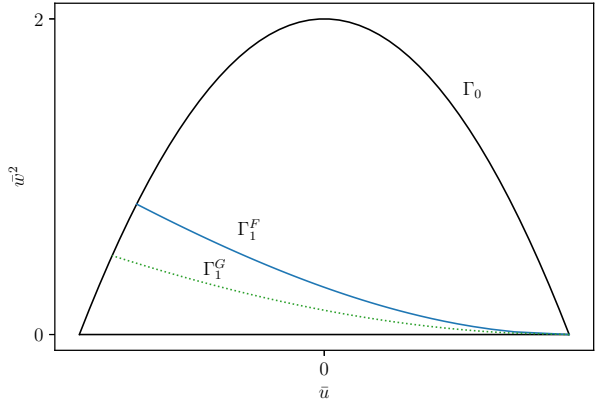}
    \caption{[Colour online] The phase space for the asymmetric wedge billiard
    if we follow the convention for the symmetric wedge billiard.}
    \label{fig:awb-phase-space-initial}
  \end{center}
\end{figure}
The lines
\begin{align}
  \Gamma_1^F &\defeq \left\{ (\bar{u}, \bar{w}^2) \in \Omega : (\bar{u}_j -
  2\bar{w}_j\cot(\theta_1))^2 \right. \notag \\
  &\qquad \left. + \bar{w}_j^2 - 2E = 0 \right\},
  \label{eqn:awb-vertex-preimage-rhs} \\
  \Gamma_1^G &\defeq \left\{ (\bar{u}, \bar{w}^2) \in \Omega : (\bar{u}_j +
  2\bar{w}_j\cot(\theta_2))^2 \right. \notag \\
  &\qquad \left. + \bar{w}_j^2 - 2E = 0 \right\}
  \label{eqn:awb-vertex-preimage-lhs}
\end{align}
are the preimages of vertex collisions for the maps $F_A$ and $G_A$
respectively.
We note that the lines coincide when $\theta_1 = \theta_2$ and that the line
$\Gamma_1^F$ lies above $\Gamma_1^G$ in the phase space $\Omega$ whenever
$\theta_1 > \theta_2$, as illustrated in Figure
\ref{fig:awb-phase-space-initial}, and vice versa.
We may suggest a division of the phase space into two or three regions possibly,
as was done for the symmetric wedge billiard; however, we note that the maps
\eqref{eqn:awb-collision-map-rhs-rhs} and \eqref{eqn:awb-collision-map-lhs-lhs}
once again map points in $\Omega$ horizontally, which might lead to a point
mapped under $F_A$ being beneath the line $\Gamma_1^G$ and thus possibly
inferred to have been mapped there by $G_A$ or possibly $F_B$.
Hence we propose that consideration should be given to a ``separation'' of the
phase space into two copies, one indicating only collisions which occur on $\p
Q_1$ and the other indicating collisions which only occur on $\p Q_2$, as
illustrated in Figure \ref{fig:awb-phase-space-separated}.
\begin{figure*}
  \begin{center}
    \includegraphics[scale=0.9]{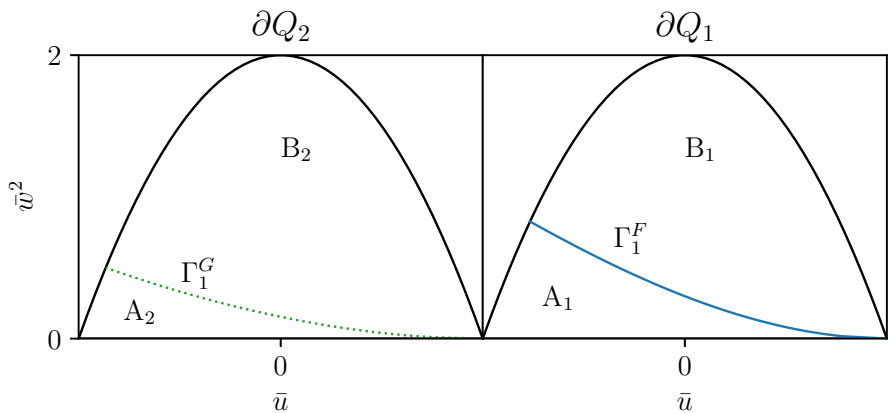}
    \caption{[Colour online] The ``separated'' phase space we propose for the
    asymmetric wedge billiard to complement the one in Figure
    \ref{fig:awb-phase-space-initial}.}
    \label{fig:awb-phase-space-separated}
  \end{center}
\end{figure*}

The region $\mathrm{A}_1$ contains points invariant under the map $F_A$ and the
region $\mathrm{A}_2$ contains points invariant under the map $G_A$.
The region $\mathrm{B}_1$ contains points mapped from $\p Q_2$ by the map $G_B$
and, similarly, the region $\mathrm{B}_2$ contains points mapped from $\p Q_1$
by the map $F_B$.
Hence the map $F_B$ maps points into either $\mathrm{A}_2$ or $\mathrm{B}_2$ and
the map $G_B$ maps points of $\p Q_2$ into either $\mathrm{A}_1$ or
$\mathrm{B}_1$.

From our simulations we noted that the case $\theta_1 + \theta_2 = \pi/2$ is
completely integrable with the phase space filled with horizontal lines, which
is similar to the dynamics of the orthogonal symmetric wedge billiard
\cite{szeredi1993classical,szeredi1996hard}.
A complete analysis of this case will be the subject of a future article by the
first author \cite{anderson2019dynamics}.
\begin{figure}
  \begin{center}
    \includegraphics{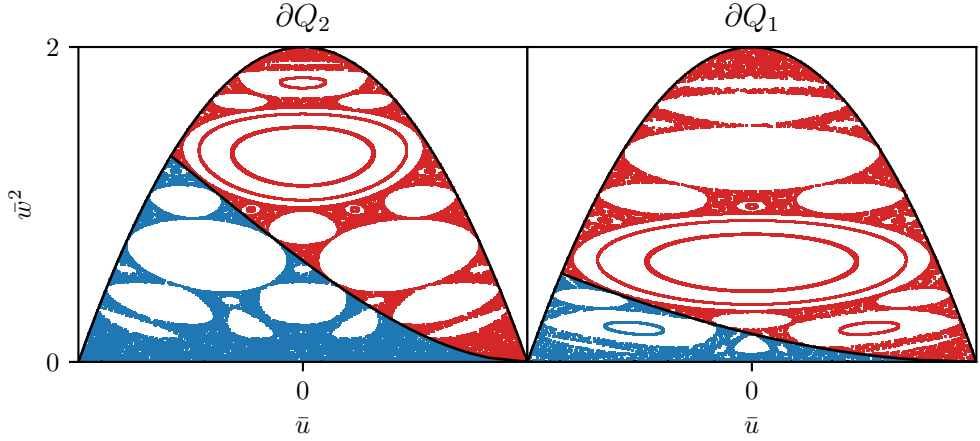}
    \caption{Phase space for $\theta_1 = 32^\circ$, $\theta_2 = 54^\circ$}
    \label{fig:awb-phase-space-35-55}
  \end{center}
\end{figure}

Furthermore, we determined that the asymmetric wedge billiard is also
completely chaotic whenever $\theta_1 + \theta_2 > \pi/2$ which agrees with the
asymmetric wedge billiard having nonvanishing Lyapunov exponents as established
by Wojtkowski \cite{wojtkowski1990system}.

For $\theta_1 + \theta_2 < \pi/2$ the behaviour once again varies between
chaotic and quasi-periodic.
However, we also noted for some parameters the phase space was completely
chaotic similar to the case of $\theta_1 + \theta_2 > \pi/2$.
We can only describe this to the broken symmetry of the asymmetric wedge and
requires further investigation.
Generally, for each fixed $\theta_1$ and $\theta_2$, the phase portraits had
points only in $\mathrm{B}_1$ and $\mathrm{B}_2$ (see Figure
\ref{fig:awb-phase-space-separated}) whenever the launch angle $\phi$ was in a
neighbourhood around $\pi/2$; this corresponds to phenomena observed in the
symmetric wedge billiard.

It was interesting to notice from our study of the phase portraits that the
asymmetric wedge billiard also bifurcated for $\theta_1 + \theta_2$ in regions
close to $\arccos((\sqrt{3} - 1)/2)$ and $\arccos((\sqrt{5} - 1)/2)$ in
correspondence with the bifurcation angles of the symmetric wedge billiard
\cite{richter1990breathing}, even though the correspondence was not exact (see
\S\ \ref{sec:awb-rotated-symmetric-wedge-billiard}).

\subsection{Rotated Symmetric Wedge Billiard}
\label{sec:awb-rotated-symmetric-wedge-billiard}
\begin{figure}
  \begin{center}
      \begin{tikzpicture}[scale=2,>=stealth]
    \coordinate (o) at (0,0);
    \draw[->] (-1.5,0) coordinate (-x) -- (2.1,0) coordinate (x);
    \draw[->] (0,-.1) -- (0,2.1) coordinate (y);
    \node at (x) [right] {$\bm{e}_1$};
    \node at (y) [above] {$\bm{e}_2$};
    \draw[thick] (o) -- ++(20:2) coordinate (a);
    \draw[thick] (o) -- ++(130:2) coordinate (b);
    \draw[dashed] (o) -- ++(75:2) coordinate (c);
    \draw (a) -- (o) -- (y) pic ["$\theta_1$", draw, angle radius=3mm, angle
    eccentricity=1.75] {angle=a--o--y};
    \draw (b) -- (o) -- (y) pic ["$\theta_2$", draw, angle radius=5mm, angle
    eccentricity=1.5] {angle=y--o--b};
    \draw (b) -- (o) -- (y) pic ["$\theta_2$", draw, angle radius=5mm, angle
    eccentricity=1.5] {angle=y--o--b};
    \path (c) -- (o) -- (y) pic ["$\varphi$", draw, angle radius=13mm, angle
    eccentricity=1.2] {angle=c--o--y};
    \fill[pattern=north east lines] (o) -- (b) -- ($(-x)!(b)!(o)$);
    \fill[pattern=north east lines] (o) -- (a) -- ($(x)!(a)!(o)$);
  \end{tikzpicture}
    \caption{The rotated symmetric wedge billiard.}
    \label{fig:awb-rotation-example}
  \end{center}
\end{figure}
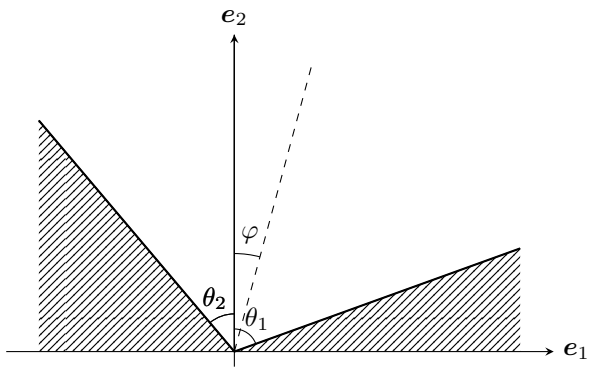

Our model enables us to consider the case of a symmetric wedge billiard with
full wedge angle rotated clockwise (or anticlockwise) from the vertical.
Let
\begin{equation}  \label{eqn:rwb-angles}
  \omega \defeq \theta_1 + \theta_2, \qquad
  \gamma \defeq \frac{\theta_1 - \theta_2}{2}
\end{equation}
be the full wedge angle and rotation angle respectively, as illustrated in
Figure \ref{fig:awb-rotation-example}.
For the rest of this section we shall assume that $\omega$ and $\gamma$ are the
given parameters.
We may solve equations \eqref{eqn:rwb-angles} for $\theta_1$, $\theta_2$ to
obtain
\begin{equation} \label{eqn:rwb-theta}
  \theta_1 = \gamma + \frac{\omega}{2}, \qquad \theta_2 = \frac{\omega}{2} -
  \gamma.
\end{equation}
Assume that we rotate the wedge clockwise, then it is more likely for $\theta_2
\to 0$ before $\theta_1 \to \pi/2$.
From the physics of the model, it follows that $0 < \theta_2 < \pi/2$ and it
follows from the second equation of \eqref{eqn:rwb-theta} that $0 < \omega/2 -
\gamma < \pi/2$ from which then follows $(\omega - \pi)/2 < \gamma <
\omega/2$.
However, $(\omega - \pi)/2 < 0$ for $\omega \in (0, \pi/2)$ and
therefore we obtain a restriction on $\gamma$ which depends on the full wedge
angle $\omega$, that is, $0 < \gamma < \omega/2$.
Hence we may not rotate the symmetric wedge further than half its full wedge
angle, which was also confirmed in our simulations.

Note that the second equation of \eqref{eqn:rwb-angles} implies that $\theta_1 >
\theta_2$ if the rotation is clockwise.
Of course, we could equally have that $\theta_1 < \theta_2$ from which would
then follow that $\gamma < 0$ which implies anticlockwise rotation from the
vertical.
In this scenario, the equations in \eqref{eqn:rwb-theta} become $\theta_1 =
\omega/2 - \gamma$ and $\theta_2 = \gamma + \omega/2$.

From our simulations of the rotated wedge billiard, we found that the dynamics
remain close to the symmetric case for very small $\gamma$.
However, as the wedge was rotated further away from the vertical, it appeared
that the phase portraits were correspondingly deformed in the vertical direction
of the phase diagrams.
As previously stated, the bifurcation angles of the symmetric wedge billiard
\cite{richter1990breathing} seem to be preserved albeit not exactly.
For example, for the bifurcation angle $\theta_1^* = \arccos((\sqrt{3} -
1)/2)/2$ rotated $\varphi = 15^\circ$ clockwise from the vertical, our
simulations indicated that the bifurcation seems to happen at $\theta_1^* =
\arccos((\sqrt{3} - 1)/2)/2 + 5/4$.
Further investigation is required to determine whether the extra term added to
$\theta_1^*$ will remain a rational number and in which way it is related to the
rotation angle $\gamma$.

\section{Conclusion}
We generalized the physical example of the wedge billiard, introduced by Lehtihet
and Miller \cite{lehtihet1986numerical} and subsequently studied by Richter
\emph{et al} \cite{richter1990breathing} and Szeredi
\cite{szeredi1993classical,szeredi1996hard} amongst others, by breaking the
symmetry of the wedge walls with respect to the vertical and considering two
separate angles $\theta_1$ and $\theta_2$ measured with respect to the vertical.

Due to the nature of the resulting nonlinear collision maps
\eqref{eqn:awb-collision-map-rhs-rhs}-\eqref{eqn:awb-collision-map-lhs-lhs}, we
undertook a computational study of the asymmetric wedge billiard and found that
the billiard is completely chaotic when $\theta_1 + \theta_2 > \pi/2$,
completely integrable when $\theta_1 + \theta_2 = \pi/2$, and varies between
quasi-periodic and chaotic motions when $\theta_1 + \theta_2 < \pi/2$.
The complete chaos observed ratifies an analytical result by Wojtkowski
\cite{wojtkowski1990system}.
%

There are some aspects which require further study.
The stability of the fixed points of $F_B$ and $G_B$ need to be determined, the
authors suspect that these fixed points are unstable for all parameter values.
There is also the matter of the bifurcation angles which are almost in exact
correspondence with the symmetric wedge billiard.
From our simulations we noted that the bifurcation occurs close to a value of
the bifurcation angle of the symmetric wedge billiard, with an added rational
number.
We suspect that there is some relationship between this rational number and the
rotation angle $\gamma$.

\bibliographystyle{abbrv}
\bibliography{refs_phd.bib}

\begin{thebibliography}{10}

\bibitem{anderson2019dynamics}
K.~D. Anderson.
\newblock Dynamics of a rotated orthogonal wedge billiard.
\newblock Unpublished.

\bibitem{anderson2019thesis}
K.~D. Anderson.
\newblock {\em Modelling and computational study of the dynamics of an
  asymmetric wedge billiard in a constant gravitational field}.
\newblock PhD thesis, University of Johannesburg, 2019.

\bibitem{berglund1996integrability}
N.~Berglund and H.~Kunz.
\newblock Integrability and ergodicity of classical billiards in a magnetic
  field.
\newblock {\em Journal of Statistical Physics}, 83(1-2):81--126, 1996.

\bibitem{berry1981regularity}
M.~V. Berry.
\newblock Regularity and chaos in classical mechanics, illustrated by three
  deformations of a circular'billiard'.
\newblock {\em European Journal of Physics}, 2(2):91, 1981.

\bibitem{birkhoff1927dynamical}
G.~D. Birkhoff.
\newblock {\em Dynamical systems}, volume~9.
\newblock American Mathematical Society, 1927.

\bibitem{birkhoff1927periodic}
G.~D. Birkhoff.
\newblock On the periodic motions of dynamical systems.
\newblock {\em Acta Mathematica}, 50(1):359--379, 1927.

\bibitem{bruus1994quantum}
H.~Bruus and A.~D. Stone.
\newblock Quantum chaos in a deformable billiard: applications to quantum dots.
\newblock {\em Physical Review B}, 50(24):18275, 1994.

\bibitem{choi2010single}
S.~Choi, B.~Sundaram, and M.~Raizen.
\newblock Single-photon cooling in a wedge billiard.
\newblock {\em Physical Review A}, 82(3):033415, 2010.

\bibitem{da2015circular}
D.~R. da~Costa, C.~P. Dettmann, and E.~D. Leonel.
\newblock Circular, elliptic and oval billiards in a gravitational field.
\newblock {\em Communications in Nonlinear Science and Numerical Simulation},
  22(1-3):731--746, 2015.

\bibitem{da2000periodic}
L.~D. Da~Silva and M.~de~Aguiar.
\newblock Periodic orbits in magnetic billiards.
\newblock {\em The European Physical Journal B: Condensed Matter and Complex
  Systems}, 16(4):719--728, 2000.

\bibitem{deryabin2003generalized1}
M.~Deryabin and L.~Pustyl'nikov.
\newblock Generalized relativistic billiards.
\newblock {\em Regular and Chaotic Dynamics}, 8(3):283--296, 2003.

\bibitem{deryabin2003generalized2}
M.~Deryabin and L.~Pustyl'nikov.
\newblock On generalized relativistic billiards in external force fields.
\newblock {\em Letters in Mathematical Physics}, 63(3):195--207, 2003.

\bibitem{deryabin2004exponential}
M.~Deryabin and L.~Pustyl’nikov.
\newblock Exponential attractors in generalized relativistic billiards.
\newblock {\em Communications in Mathematical Physics}, 248(3):527--552, 2004.

\bibitem{gongora2002classical}
A.~G{\'o}ngora-T, J.~V. Jos{\'e}, and S.~Schaffner.
\newblock Classical solutions of an electron in magnetized wedge billiards.
\newblock {\em Physical Review E}, 66(4):047201, 2002.

\bibitem{greene1981universal}
J.~M. Greene, R.~MacKay, F.~Vivaldi, and M.~Feigenbaum.
\newblock Universal behaviour in families of area-preserving maps.
\newblock {\em Physica D: Nonlinear Phenomena}, 3(3):468--486, 1981.

\bibitem{hale1991dynamics}
J.~H. Hale and H.~Kocak.
\newblock {\em Dynamics and Bifurcations}.
\newblock Spring-Verlag, 1991.

\bibitem{hartl2013dynamics}
A.~Hartl, B.~Miller, and P.~Mazzoleni.
\newblock Dynamics of a dissipative, inelastic gravitational billiard.
\newblock {\em Physical Review E}, 87:032901, 2013.

\bibitem{kamphorst1999bounded}
S.~Kamphorst and S.~De~Carvalho.
\newblock Bounded gain of energy on the breathing circle billiard.
\newblock {\em Nonlinearity}, 12, 1999.

\bibitem{koiller1995time}
J.~Koiller, R.~Markarian, S.~Kamphorst, and S.~De~Carvalho.
\newblock Time-dependent billiards.
\newblock {\em Nonlinearity}, 8, 1995.

\bibitem{korsch1991new}
H.~Korsch and J.~Lang.
\newblock A new integrable gravitational billiard.
\newblock {\em Journal of Physics A: Mathematical and General}, 24(1):45, 1991.

\bibitem{ladeira2008scaling}
D.~G. Ladeira and J.~K.~L. da~Silva.
\newblock Scaling features of a breathing circular billiard.
\newblock {\em Journal of Physics A: Mathematical and Theoretical},
  41(36):365101, 2008.

\bibitem{lazutkin1973existence}
V.~F. Lazutkin.
\newblock The existence of caustics for a billiard problem in a convex domain.
\newblock {\em Mathematics of the USSR-Izvestiya}, 7(1):185, 1973.

\bibitem{lehtihet1986numerical}
H.~Lehtihet and B.~Miller.
\newblock Numerical study of a billiard in a gravitational field.
\newblock {\em Physica D: Nonlinear Phenomena}, 21(1):93--104, 1986.

\bibitem{lenz2007classical}
F.~Lenz, F.~K. Diakonos, and P.~Schmelcher.
\newblock Classical dynamics of the time-dependent elliptical billiard.
\newblock {\em Physical Review E}, 76(6):066213, 2007.

\bibitem{lenz2007scattering}
F.~Lenz, F.~K. Diakonos, and P.~Schmelcher.
\newblock Scattering dynamics of driven closed billiards.
\newblock {\em EPL (Europhysics Letters)}, 79(2):20002, 2007.

\bibitem{lenz2009evolutionary}
F.~Lenz, C.~Petri, F.~Koch, F.~Diakonos, and P.~Schmelcher.
\newblock Evolutionary phase space in driven elliptical billiards.
\newblock {\em New Journal of Physics}, 11(8):083035, 2009.

\bibitem{pina1987symmetry}
E.~Pi{\~n}a and L.~J. Lara.
\newblock On the symmetry lines of the standard mapping.
\newblock {\em Physica D: Nonlinear Phenomena}, 26(1-3):369--378, 1987.

\bibitem{poritsky1950billiard}
H.~Poritsky.
\newblock The billard ball problem on a table with a convex boundary--an
  illustrative dynamical problem.
\newblock {\em Annals of Mathematics}, 51(2):pp. 446--470, 1950.

\bibitem{richter1990breathing}
P.~H. Richter, H.-J. Scholz, and A.~Wittek.
\newblock A breathing chaos.
\newblock {\em Nonlinearity}, 3(1):45, 1990.

\bibitem{robnik1985classical}
M.~Robnik and M.~V. Berry.
\newblock Classical billiards in magnetic fields.
\newblock {\em Journal of Physics A: Mathematical and General},
  18(9):1361--1378, 1985.

\bibitem{sepulchre2003stabilization}
R.~Sepulchre and M.~Gerard.
\newblock Stabilization of periodic orbits in a wedge billiard.
\newblock In {\em 42nd IEEE International Conference on Decision and Control
  (IEEE Cat. No. 03CH37475)}, volume~2, pages 1568--1573. IEEE, 2003.

\bibitem{sinai1970dynamical}
Y.~G. Sinai.
\newblock Dynamical systems with elastic reflections: ergodic properties of
  dispersing billiards.
\newblock {\em Uspekhi Matematicheskikh Nauk}, 25(2):141--192, 1970.

\bibitem{szeredi1993classical}
T.~Szeredi.
\newblock {\em Classical and quantum chaos in the wedge billiard}.
\newblock PhD thesis, McMaster University, 1993.

\bibitem{szeredi1996hard}
T.~Szeredi.
\newblock Hard chaos and adiabatic quantization: The wedge billiard.
\newblock {\em Journal of Statistical Physics}, 83(1-2):259--274, 1996.

\bibitem{szeredi1992periodic}
T.~Szeredi and D.~Goodings.
\newblock Periodic orbits from the quantum energy spectrum of the wedge
  billiard.
\newblock {\em Physical Review Letters}, 69(11):1640, 1992.

\bibitem{szeredi1993classical1}
T.~Szeredi and D.~Goodings.
\newblock Classical and quantum chaos of the wedge billiard. i. classical
  mechanics.
\newblock {\em Physical Review E}, 48(5):3518, 1993.

\bibitem{szeredi1993classical2}
T.~Szeredi and D.~Goodings.
\newblock Classical and quantum chaos of the wedge billiard. ii. quantum
  mechanics and quantization rules.
\newblock {\em Physical Review E}, 48(5):3529, 1993.

\bibitem{tasnadi1996behavior}
T.~Tasn{\'a}di.
\newblock The behavior of nearby trajectories in magnetic billiards.
\newblock {\em Journal of Mathematical Physics}, 37(11):5577--5598, 1996.

\bibitem{waalkens1997elliptic}
H.~Waalkens, J.~Wiersig, and H.~R. Dullin.
\newblock Elliptic quantum billiard.
\newblock {\em Annals of Physics}, 260(1):50--90, 1997.

\bibitem{wojtkowski1986principles}
M.~Wojtkowski.
\newblock Principles for the design of billiards with nonvanishing lyapunov
  exponents.
\newblock {\em Communications in Mathematical Physics}, 105(3):391--414, 1986.

\bibitem{wojtkowski1990system}
M.~P. Wojtkowski.
\newblock A system of one dimensional balls with gravity.
\newblock {\em Communications in Mathematical Physics}, 126(3):507--533, 1990.

\end{thebibliography}

\end{document}